\documentclass[twocolumn,superscriptaddress,nobibnotes,nofootinbib,floatfix,aps,prl,citeautoscript,reprint]{revtex4-1}

\usepackage{graphicx}
\usepackage{epsfig}
\usepackage{subfigure}
\usepackage{color}
\usepackage{latexsym}
\usepackage{amsmath,bm,multirow}
\usepackage{amsfonts}
\usepackage{dcolumn}           
\usepackage{natbib}
\usepackage[colorlinks,linkcolor=blue,citecolor=green]{hyperref}
\usepackage{physics}

\usepackage[draft]{changes} 
\usepackage[normalem]{ulem}
\definechangesauthor[name={yi}, color=blue]{yi}
\newcommand{\argonne}{Center for Nanoscale Materials, Argonne National Laboratory, Argonne, IL 60439, USA}

\usepackage[draft]{changes} 
\usepackage[normalem]{ulem}
\definechangesauthor[name={Maria}, color=red]{Maria}
\definechangesauthor[name={Yi}, color=blue]{Yi}
\usepackage{soul}

\begin{document}

\title{Renormalized Lattice Dynamics and Thermal Transport in VO$_{2}$}

\author{Yi Xia} 
\email{yxia@anl.gov}
\affiliation{\argonne}
\author{Maria K. Y. Chan}
\email{mchan@anl.gov}
\affiliation{\argonne}

\date{\today}

\begin{abstract}
Vanadium dioxide (VO$_{2}$) undergoes a first-order metal-insulator transition (MIT) upon cooling near room temperature, concomitant with structural change from rutile to monoclinic. Accurate characterization of lattice vibrations is vital for elucidating the transition mechanism. To investigate the lattice dynamics and thermal transport properties of VO$_{2}$ across the MIT, we present a phonon renormalization scheme based on self-consistent phonon theory through iteratively refining vibrational free energy. Using this technique, we compute temperature-dependent phonon dispersion and lifetimes, and point out the importance of both magnetic and vibrational entropy in driving the MIT. The predicted phonon dispersion and lifetimes show quantitative agreement with experimental measurements. We demonstrate that lattice thermal conductivity of rutile VO$_{2}$ is nearly temperature independent as a result of strong intrinsic anharmonicity, while that of monoclinic VO$_{2}$ varies according to $1/T$. Due to phonon softening and enhanced scattering rates, the lattice thermal conductivity is deduced to be substantially lower in the rutile phase, suggesting that Wiedemann-Franz law might not be strongly violated in rutile VO$_{2}$.
\end{abstract}

\maketitle


Vanadium dioxide (VO$_{2}$) is a transition metal oxide under intense study, because it exhibits a first-order metal-insulator transition (MIT) from metallic rutile phase (R-VO$_{2}$) to semiconducting monoclinic phase (M-VO$_{2}$) upon cooling to $T_{\text{MIT}}$=340\ K, which facilitates near-room-temperature switching of either electronic or thermal conductivity \cite{Pergament2017}. The mechanism underlying the MIT in VO$_{2}$ has been a longstanding subject of controversy due to the coupled changes in structural and electronic properties across the MIT. Various studies have focused on clarifying whether the MIT is driven by instabilities in electron-lattice dynamics or charge localization due to strong electron-electron correlation \cite{Wentzcovitch1994,Hearn1972,Zylbersztejn1975,Eyert2002,Goodenough1971}. A recent study \cite{Budai2014} suggested the importance of vibrational entropy in driving the MIT based on X-ray/neutron scattering measurements and \textit{ab initio} molecular dynamics (AIMD) calculations \cite{Hellman2013}. By examining experimentally-measured temperature-dependent phonon density of states (DOS), a discontinuity in DOS shape across the MIT is identified, which is associated with the softening of transverse acoustic modes. The change in vibrational entropy across the MIT was found to be $\Delta S_{\text{ph}}$=1.02$\pm$0.09 $k_{\text{B}}$/VO$_{2}$, which is comparable to the total entropy change of 1.50$\pm$0.15 $k_{\text{B}}$/VO$_{2}$ from latent heat measurements \cite{Berglund1969}, suggesting that phonons dominate over electrons to stabilize R-VO$_{2}$. More recently however, quantum Monte Carlo (QMC) calculations \cite{Huihuo2015} show that while both phases have antiferromagnetic ground states, a significant change in magnetic coupling strength across the transition leads to magnetic ordering and disordering in M- and R-VO$_{2}$ respectively, thus contributing additional entropy to stabilize R-VO$_{2}$ \cite{Huihuo2016}. Accurately determining phonon entropy change is therefore important towards resolving the transition mechanism.

While electrical conductivity increase by several orders of magnitude is well documented, only a few studies address the separate contributions from electrons and phonons to thermal transport in VO$_{2}$ across the MIT, primarily due to coupled electron and phonon transport in metallic R-VO$_{2}$. In a recent study, \citeauthor{Lee371} \cite{Lee371} attempted to decouple the thermal conductivity of R-VO$_{2}$ into electron ($\kappa_{e}$) and phonon ($\kappa_{l}$) contributions, by deducing $\kappa_{l}$ using experimentally-measured phonon linewidths and computed phonon dispersions. The resultant value of $\kappa_{e}$ in R-VO$_{2}$ was anomalously low, implying a significant violation of the Wiedemann-Franz (WF) law, which the authors attributed to unconventional quasiparticle dynamics, i.e. absence of quasiparticles in a strongly-correlated electron fluid where heat and charge diffuse independently \cite{Lee371}.

Since $\kappa_{e}$ can only be indirectly determined experimentally, by subtracting $\kappa_{l}$ from the total thermal conductivity measurements, accurate modeling of $\kappa_{l}$ of R-VO$_{2}$ is crucial. From a theoretical point of view, R-VO$_{2}$ is challenging due to its strong anharmonicity and lattice metastability. We present a phonon renormalization scheme based on iteratively refining vibrational free energy, temperature-dependent interatomic force constants (IFCs) obtained using compressive sensing lattice dynamics (CSLD) \cite{csld}; and application of this scheme towards a comprehensive study of lattice dynamics and thermal transport properties of VO$_{2}$. Our study quantitatively clarifies the the changes in vibrational entropy and $\kappa_{l}$ across the MIT, demonstrates the importance of magnetic entropy, and suggests that WF law may not be strongly violated in R-VO$_{2}$.


Existing methods that are capable of treating strong anharmonic effects nonperturbatively are based on or related to self-consistent phonon theory \cite{Werthamer1970}, such as self-consistent \textit{ab initio} lattice dynamics (SCAILD) \cite{Souvatzis2009} and stochastic self-consistent harmonic approximation (SSCHA) \cite{Errea2014}. The present implementation of phonon renormalization is in the spirit of SCAILD, where vibrational free energy is iteratively refined based on temperature-dependent atomic displacements. As implemented by Roekeghem \textit{et.\ al.} \cite{Roekeghem2016}, an improvement over SCAILD can be achieved by using the full quantum mean square thermal displacement matrix, which allows for simultaneous update of both phonon frequencies and eigenvectors. The temperature-dependent atomic displacements $\{u_{a,\alpha}\}$ are generated according to the probability $\rho (\{u_{a,\alpha}\}) \propto \text{exp}( -\frac{1}{2} \mathbf{u} \Sigma^{-1} \mathbf{u} )$ to find a displaced configuration in the harmonic approximation, where $u_{a,\alpha}$ is the displacement of atom $a$ in $\alpha$ direction, and $\Sigma$ is known as the quantum covariance matrix of displacement vector \cite{Errea2014,Roekeghem2016}
	\begin{equation}\label{eq:covar}
	\Sigma_{a\alpha, b\beta} = \frac{\hbar}{2\sqrt{M_{a}M_{b}}} \sum_{\lambda} \frac{\left(1+2n_{\omega_{\lambda}}\right) }{\omega_{\lambda}} \epsilon^{\lambda}_{a\alpha} \epsilon^{\ast\lambda}_{b\beta},
	\end{equation}
where $M$, $\omega$, $\epsilon$, $n$, $\lambda$ are atomic mass, phonon frequency, eigenvector, Bose-Einstein distribution and branch index respectively. Various sets of $\{u_{a,\alpha}\}$ can be generated using the covariance matrix following a given distribution, such as continuous Gaussian distribution or discrete Rademacher distribution. Note that the current scheme to generate temperature-dependent atomic displacements is superior to sampling trajectory from AIMD, which suffers from the absence of nuclear quantum effects and thus underestimates the thermal displacements for temperatures below the Debye temperature ($\approx$ 1000 K for VO$_{2}$). A self-consistent loop was formed by generating $\{u_{a,\alpha}\}$ using effective IFCs extracted from the previous step. Both $2^{\text{nd}}$- and $3^{\text{rd}}$-order IFCs were simultaneously extracted using a recently developed method named compressive sensing lattice dynamics (CSLD) \cite{csld}, wherein DFT forces are expressed as a Taylor expansion in displacements and the coefficients are obtained from sparse regression. From the extracted effective IFCs, phonon lifetimes were evaluated using Fermi's golden rule by treating 3$^{\text{rd}}$-order IFCs as perturbation to harmonic phonons \cite{ziman}, and linearized Boltzmann transport equation (BTE) was solved in an iterative manner to account for the non-equilibrium phonon distributions \cite{omini1,omini2,broido,wuli}. (See Supplemental Material \cite{VO2SI} for more details of CSLD and implementation/validation of our renormalization scheme, and  structural and computational details.)

\begin{figure*}[htp]
	\includegraphics[width = 0.95\linewidth]{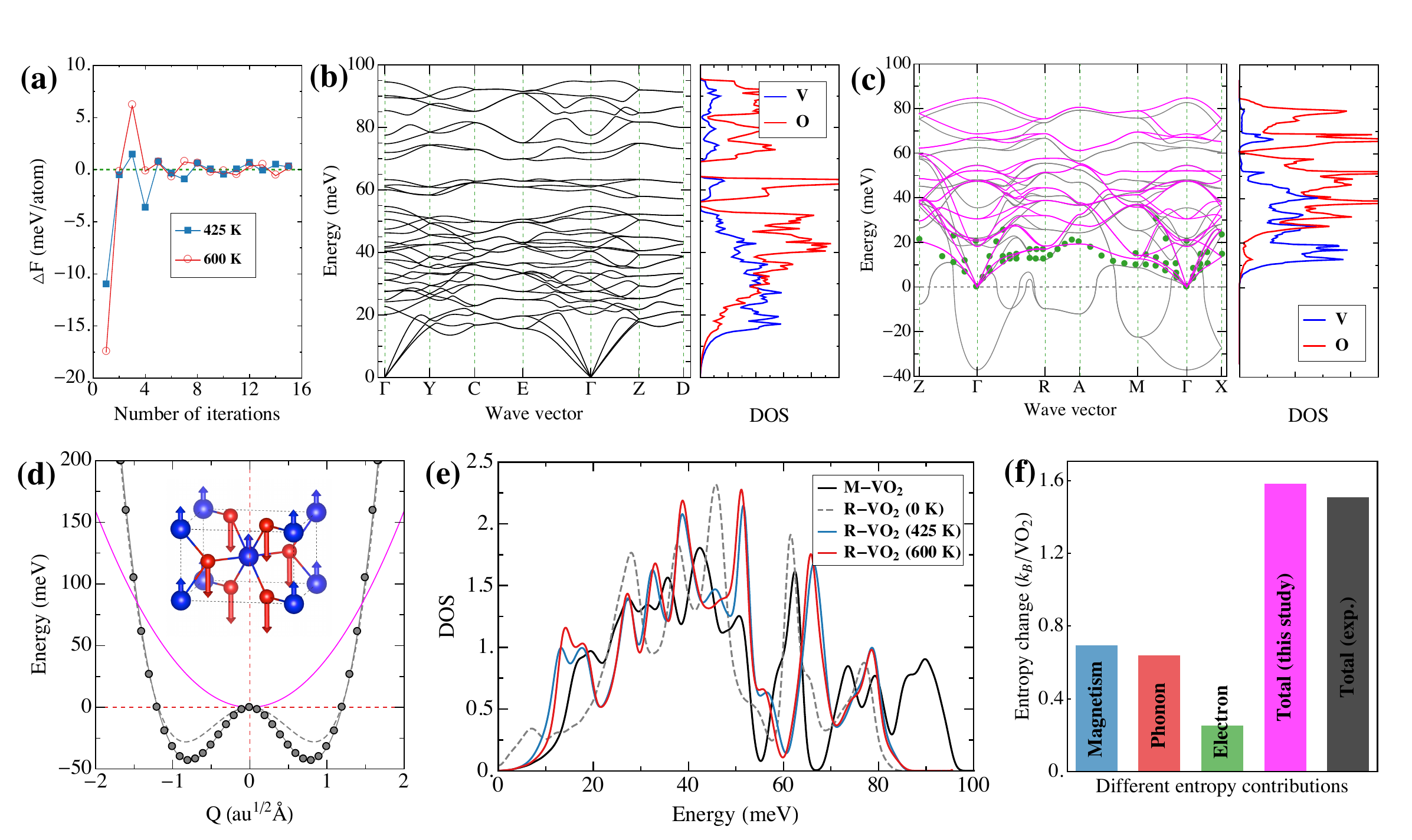}
	\caption{ 
	(a) The change of vibrational free energy as a function of iterative steps in the phonon renormalization process of R-VO$_{2}$ at 425 and 600 K. The green dashed line denotes the converged value of vibrational free energy (set to zero). (b) Computed phonon dispersions and atom-projected DOS of M-VO$_{2}$ at 0 K. (c) Renormalized phonon dispersions (solid magenta lines) and atom-projected DOS of R-VO$_{2}$ at 425 K in comparison with experimental IXS data at 425 K \cite{Budai2014} (green disks) and the phonon dispersions computed at 0 K (solid gray lines). Imaginary frequencies are denoted as negative numbers. (d) Potential energy surface of the zone-center imaginary phonon mode of R-VO$_{2}$ as a function of vibrational amplitude. The solid gray disks denote the computed energies, and dashed/solid gray line denotes polynomial fitting up to 6th/16th order. The solid magenta line represents the effective harmonic potential at 425 K. The inset shows the atomic displacements in accordance with the eigenvector. (e) Total phonon DOS of M-VO$_{2}$ at 0 K (solid black lines) and R-VO$_{2}$ at 0 K (dashed gray lines), 425 K (solid blue lines), and 600 K (solid red lines). (f) Separate entropy contributions from magnetic ordering \cite{Huihuo2016}, phonons, and electrons \cite{Budai2014} across the metal-insulator transition, compared with total entropy change measured from latent heat experiment \cite{Berglund1969}.
	}.
	\label{fig:Disp_DOS}
\end{figure*}


To confirm the convergence of renormalized phonon dispersions, vibrational free energy of R-VO$_{2}$ computed in each iteration, as shown in Fig.~\ref{fig:Disp_DOS}(a), is examined and found to achieve good convergence to about $\pm$1 meV/atom after seven iterations. Without accounting for temperature effects, phonon dispersions of M-VO$_{2}$ computed in the harmonic approximation, as shown in Fig.~\ref{fig:Disp_DOS}(b), have well defined frequencies with real values, exhibiting normal harmonic behavior. In contrast, there is a large number of imaginary frequencies in phonon dispersions of R-VO$_{2}$ across the Brillouin zone, consistent with previous theoretical studies \cite{Kim2013,Budai2014,Lee371}. With temperature effects considered, the renormalized phonon dispersions of R-VO$_{2}$ at 425 K in Fig.~\ref{fig:Disp_DOS}(c) exhibit hardening of optical modes and achieve overall good agreement with experimental IXS measurements \cite{Budai2014}, confirming the validity of the phonon renormalization approach. Specifically, the computed high-lying transverse acoustic (TA2), longitudinal acoustic (LA) and low-lying zone-center optical (ZCO) phonon frequencies are in excellent agreement with experiments, while low-lying transverse acoustic (TA1) phonon frequencies display some overestimation, particularly along $\Gamma$-R and $\Gamma$-M paths. Renormalized phonon dispersions of R-VO$_{2}$ were additionally computed at 360 and 600 K to investigate the temperature dependence. As shown in Fig.~S6 \cite{VO2SI}, the presence of a few imaginary frequencies at 360 K indicates that R-VO$_{2}$ tends to be unstable near MIT. Only the low-lying ZCO phonon mode is significantly hardened (from 17.8 meV to 23.3 meV) with temperature enhanced from 425 to 600 K, indicating that the ZCO phonon mode is severely anharmonic. 

To quantitatively investigate the anharmonicity of the ZCO phonon mode, Fig.~\ref{fig:Disp_DOS}(d) shows its potential energy surface as a function of atomic displacement following the renormalized eigenvector, wherein V and O atoms move away from each other along the rutile $c$ axis. The resultant double-well potential of the ZCO phonon mode demonstrates the structural instability of R-VO$_{2}$, which appears in 0 K phonon dispersions. Indeed, by projecting the renormalized eigenvector onto those zone-center phonon modes computed at 0 K, it is found that ZCO phonon mode matches exactly the zone-center imaginary optical mode. Therefore, the structure of R-VO$_{2}$ is a dynamic average over symmetry-broken minima separated by relatively deep energetic barriers ($\approx$ 42 meV), and the positions of V and O atoms represent their averaged spatial occupations, consistent with experimental observations of large thermal displacements \cite{McWhan1974}. The severe anharmonicity of the ZCO mode is also reflected in the detailed shape of the energy profile, an accurate fitting of which requiring polynomials well beyond 6$^\text{th}$-order. 

To confirm and explain the experimentally observed large phonon entropy change across the MIT \cite{Budai2014}, the total phonon DOS of M- and R-VO$_{2}$ are compared in Fig.~\ref{fig:Disp_DOS}(e). It can be seen that temperature-induced phonon renormalization significantly alters the DOS of R-VO$_{2}$. Further increasing temperature from 425 K to 600 K tends to slightly harden low-lying modes while softening high-lying modes. Comparison of total DOS between M- and R-VO$_{2}$ displays an abrupt blueshift of low-lying DOS shoulders ($\approx$ 13 meV) with phonon modes notably hardened when temperature is reduced below the MIT. These phonon modes are mainly associated with the TA1 modes near zone boundary and constitute the majority of vibrational entropy change. To quantify it, temperature-dependent vibrational entropy is computed using phonon dispersions of M-VO$_{2}$ at 0 K and R-VO$_{2}$ at 425 K. As shown in Fig.~\ref{fig:Disp_DOS}(f), both PBE and PBEsol xc functionals are found to yield similar vibrational entropy change of 0.64 $k_{\text{B}}$/VO$_{2}$ at 340 K, which is smaller than 1.02 $k_{\text{B}}$/VO$_{2}$ reported by \textcite{Budai2014}. As a result, total entropy change, which includes contributions of 0.64 and 0.25 $k_{\text{B}}$/VO$_{2}$ from lattice vibration and partial occupancy of electrons \cite{Budai2014} in R-VO$_{2}$ respectively, is smaller than experimental report of 1.50 $k_{\text{B}}$/VO$_{2}$ \cite{Berglund1969}. The additional entropy change may be explained by the magnetic contribution revealed by the aforementioned QMC study \cite{Huihuo2015,Huihuo2016}. In M-VO$_{2}$, singlets are formed due to the strongly coupled magnetic moments within the dimers, which leads to zero entropy for such a state. Whereas above the transition temperature, R-VO$_{2}$ is in a magnetic disordered state, leading to an entropy change of $\Delta S = k_{\text{B}}\text{ln}(2) = 0.69$  $k_{\text{B}}$/VO$_{2}$ across the MIT. As shown in Fig.~\ref{fig:Disp_DOS}(f), the total entropy change accounting for electron, phonon and magnetic contributions is 1.58 $k_{\text{B}}$/VO$_{2}$, which is in excellent agreement with experimental value of 1.50 $k_{\text{B}}$/VO$_{2}$ \cite{Berglund1969}, implying that magnetic ordering is also important in stabilizing the rutile phase.

\begin{figure*}[htp]
	\includegraphics[width = 1.0\linewidth]{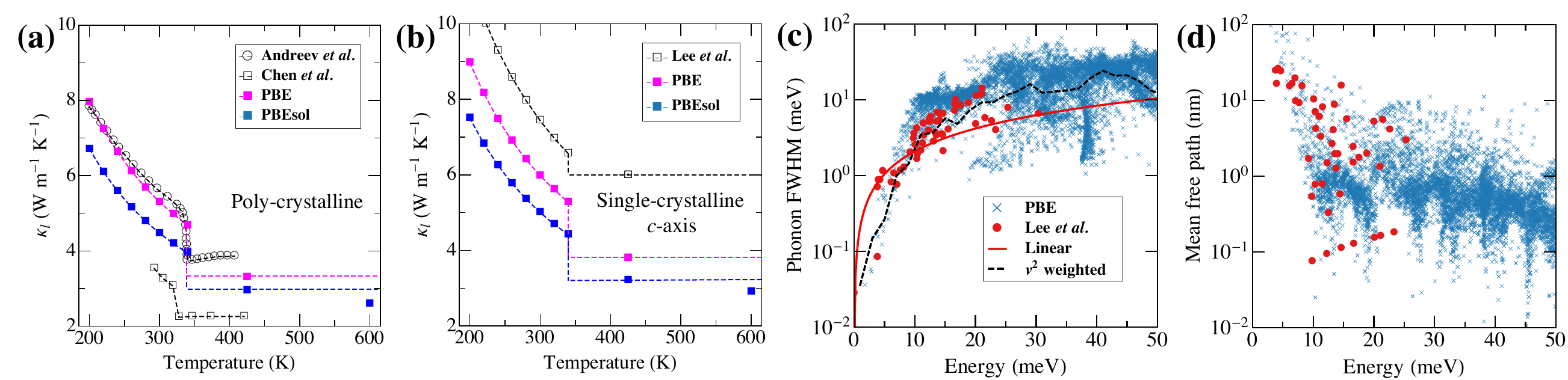}
	\caption{
	 (a) Computed $\kappa_{l}$ of polycrystalline VO$_{2}$ using PBE and PBEsol exchange correlation functionals in comparison with experimental measurements reported by \textcite{ANDREEV1978} and \textcite{Chen2012}. (b) Computed $\kappa_{l}$ of single-crystal VO$_{2}$ along the rutile $c$ axis in comparison with results by \textcite{Lee371}. (c) and (d) Computed phonon mode FWHM and MFP (blue crosses) using PBE exchange correlation functional in comparison with experimental measurements (red disks) at 425 K \cite{Lee371}. The solid red line denotes the linear relation between FWHM and phonon energy employed in Ref.\ [\onlinecite{Lee371}]. The dashed black line denotes the group-velocity-weighted FWHM averaged over phonon modes with similar energies.
	 }
	\label{fig:Kappa_Ana}
\end{figure*}

Having validated the present phonon renormalization approach by comparing the computed phonon dispersions and vibrational entropy change against experiments, we calculate lattice thermal transport properties and compare them with experiments in Fig.~\ref{fig:Kappa_Ana}. Considering that experiments found that both phonon lifetimes and $\kappa_{l}$ of R-VO$_{2}$ show no temperature dependence up to 425 K \cite{Budai2014,Lee371}, the $\kappa_{l}$ of R-VO$_{2}$ was computed using renormalized IFCs at 425 K. By comparing the computed $\kappa_{l}$ of polycrystalline VO$_{2}$ to experimental measurements reported by \textcite{ANDREEV1978} and \textcite{Chen2012}, as shown in Fig.~\ref{fig:Kappa_Ana}(a), it is found that both theoretical and experimental results show $1/T$ dependence of $\kappa_{l}$ in M-VO$_{2}$. The $\kappa_{l}$ of R-VO$_{2}$ at 600 K was further evaluated to investigate its temperature dependence. The PBEsol results show a slightly decreased $\kappa_{l}$ from 2.95 to 2.60 Wm$^{-1}$K$^{-1}$ with temperature increased from 425 to 600 K, which is considerably slower than $1/T$ and consistent with the ``amorphous" character of heat conduction without presence of significant structural disorder found in Ref.~\cite{ANDREEV1978}. We find that this weak temperature dependence of $\kappa_{l}$ is due to strong anharmonicity, which significantly renormalizes 2$^{\text{nd}}$ and 3$^{\text{rd}}$-order IFCs.

Compared to experiments, PBE results agree very well with Ref.~[\onlinecite{ANDREEV1978}], while both PBE and PBEsol results exhibit overestimation compared to Ref.~[\onlinecite{Chen2012}], the later of which is probably caused by the presence of defects such as grain boundaries and pores in R-VO$_{2}$ \cite{Chen2012}. Nevertheless, both PBE and PBEsol yield significant decrease of $\kappa_{l}$ across the MIT (1.4 and 1.0 Wm$^{-1}$K$^{-1}$ for PBE and PBEsol respectively), which is comparable with the value of 1.3 Wm$^{-1}$K$^{-1}$ reported in Ref.~[\onlinecite{ANDREEV1978}]. Since $\kappa_{l}$ of R-VO$_{2}$ in Ref.~[\onlinecite{ANDREEV1978}] is evaluated using standard Lorentz number $L_{0}$, the present theoretical results suggest that WF law is not strongly violated in polycrystalline VO$_{2}$. However, compared to the estimated $\kappa_{l}$ along the rutile $c$ axis using first-principles phonon dispersions and phonon scattering rates obtained from the IXS measurements \cite{Lee371} as shown in Fig.~\ref{fig:Kappa_Ana}(b), both PBE and PBEsol results display smaller $\kappa_{l}$ for M- and R-VO$_{2}$ and a much larger decrease of $\kappa_{l}$ across the MIT, which is primarily due to significant differences in the $\kappa_{l}$ of R-VO$_{2}$. Since $\kappa_{l}$ is subtracted from total thermal conductivity to estimate $\kappa_{e}$ in Ref.~\cite{Lee371}, which is subsequently utilized to identify a strong violation of WF law in R-VO$_{2}$, it is crucial to confirm and verify $\kappa_{l}$ of R-VO$_{2}$. 

To explore the origin of our predicted lower $\kappa_{l}$ of R-VO$_2$, the full widths at half maximum (FWHM = $h / \tau$ with lifetime $\tau$) and mean free paths (MFP = $|\mathbf{v}| \tau$, where $\mathbf{v}$ is group velocity) of phonon modes at 425 K and varying energies are compared with those estimated using IXS measurements \cite{Lee371}. As shown in Fig.~\ref{fig:Kappa_Ana}(c), the predicted FWHMs are consistent with IXS data. Considering the fact that there are many different lifetimes for phonon modes with similar energies and IXS samples a limited space in the full Brillouin zone, a better comparison with experiments can be achieved by computing energy-dependent FWHM weighted by squared phonon group velocity, which is a more rigorous way to average the FWHM than linearly interpolating the phonon modes with small FWHMs as adopted in Ref.~[\onlinecite{Lee371}]. Our predicted energy-dependent FWHM agrees quantitatively with IXS data from low to high phonon energies, further verifying the accuracy of computed phonon scattering rates. The same good agreement is also achieved in mode-resolved MFP, as shown in Fig.~\ref{fig:Kappa_Ana}(d). Our results imply that simple linear interpolation tends to significantly overestimate/underestimate FWHMs of phonon modes with low/high energies, in addition to neglecting the strong anisotropy of mode-dependent FWHM. Since the energy cumulative $\kappa_{l}$ (see Fig.\ S7 \cite{VO2SI}) shows that phonon modes with energies larger than 10 meV contribute about 70\% of total $\kappa_{l}$, the $\kappa_{l}$ of R-VO$_{2}$ deduced by linearly interpolated FWHMs in Ref.\ [\onlinecite{Lee371}] is therefore overestimated. 

Note that the present predicted $\kappa_{l}$ of R-VO$_{2}$ is also potentially overestimated because (1) the renormalized TA1 modes have slightly higher energies than experimental measurements, which commonly lead to larger $\kappa_{l}$ because of reduced scattering rates and increased group velocity, and (2) high-order phonon-phonon interactions beyond three phonon processes are neglected in our calculations. To illustrate the drop of $\kappa_{l}$ across the MIT upon heating, comparisons of mode-resolved phonon lifetime, group velocity and MFP between M- and R-VO$_{2}$ are shown in Fig.\ S8(a)-(c) \cite{VO2SI}. The phonon lifetimes of M-VO$_{2}$ are overall significantly longer than those of R-VO$_{2}$, while the phonon group velocities of both R- and M-VO$_{2}$ are fairly similar, resulting overall larger MFPs in M-VO$_{2}$ than in R-VO$_{2}$. As a consequence, a substantial change of $\kappa_{l}$ across the MIT is expected, suggesting that WF law may not be strongly violated as demonstrated in Ref.~\ [\onlinecite{Lee371}]. 

In summary, we have applied a first-principles-based phonon renormalization scheme to model lattice dynamics and thermal transport properties of VO$_{2}$ across the metal-insulator transition. Using the obtained temperature-dependent 2$^{\text{nd}}$- and 3$^{\text{rd}}$-order interatomic force constants, we computed renormalized phonon dispersions of rutile VO$_{2}$ and identified its intrinsic strong anharmonicity associated with low-lying zone-center optical mode. We also computed vibrational entropy change across the MIT, based on which we found that contribution from magnetic entropy may also be important. Finally, lattice thermal transport properties were investigated for both phases and a significant change of $\kappa_{l}$ across the MIT is confirmed, suggesting that Wiedemann-Franz law might not be strongly violated in R-VO$_{2}$.


\textbf{Acknowledgements} We thank Huihuo Zheng for fruitful discussions. This material is based upon work supported by Laboratory Directed Research and Development (LDRD) funding from Argonne National Laboratory, provided by the Director, Office of Science, of the U.S. Department of Energy under Contract No. DE-AC02-06CH11357. Use of the Center for Nanoscale Materials, an Office of Science user facility, was supported by the U.S. Department of Energy, Office of Science, Office of Basic Energy Sciences, under Contract No. DE-AC02-06CH11357. This research used resources of the National Energy Research Scientific Computing Center, a DOE Office of Science User Facility supported by the Office of Science of the U.S. Department of Energy under Contract No. DE-AC02-05CH11231.

\bibliography{VO2}

\widetext
\clearpage
\begin{center}
	\textbf{\large Supplementary Materials: Renormalized Lattice Dynamics and Thermal Transport in VO$_{2}$}
\end{center}

\begin{center}

Yi Xia$^{1}$ and  Maria K. Y. Chan$^{1}$

\vspace{0.3cm}

\text{$^1$\argonne}

\end{center}

\tableofcontents

\setcounter{equation}{0}
\setcounter{figure}{0}
\setcounter{table}{0}
\setcounter{page}{1}
\makeatletter
\renewcommand{\thepage}{S\arabic{page}}
\renewcommand{\theequation}{S\arabic{equation}}
\renewcommand{\thefigure}{S\arabic{figure}}
\renewcommand{\thetable}{S\arabic{table}}
\renewcommand{\bibnumfmt}[1]{[S#1]}
\renewcommand{\citenumfont}[1]{S#1}



\newpage
\section{Methodology}

\subsection{Compressive sensing lattice dynamics}
Compressive sensing lattice dynamics (CSLD) \cite{csld} was used to extract the 2$^\text{nd}$- and 3$^\text{rd}$-order interatomic force constants (IFCs). Within CSLD, the total energy is expressed as Taylor expansion in terms of the atomic displacements
\begin{equation}
\label{eq:Taylor}
	E=E_{0}+\Phi_{\mathbf{a}}u_{\mathbf{a}}+\frac{1}{2}\Phi_{\mathbf{ab}}u_{\mathbf{a}}u_{\mathbf{b}}+\frac{1}{3!}\Phi_{\mathbf{abc}}u_{\mathbf{a}}u_{\mathbf{b}}u_{\mathbf{c}}+\cdots,
\end{equation}
where $E_{0}$ is the ground state energy, $u_{\mathbf{a}} \equiv u_{a,\alpha}$ is the displacement of atom $a$ in the Cartesian direction $\alpha$, and $\Phi_\mathbf{ab}$, $\Phi_{\mathbf{abc}}$ are the harmonic and 3$^\text{rd}$-order IFCs. The linear term with $\Phi_\mathbf{a}$ is absent if the reference lattice sites represent mechanical equilibrium, and the Einstein summation convention over repeated indices is used. CSLD belongs to the class of direct supercell methods where DFT forces are used to ``invert'' Eq.~\eqref{eq:Taylor} using the force-displacement relation
\begin{equation}
	F_{\mathbf{a}}=-\Phi_{\mathbf{a}}-\Phi_{\mathbf{ab}}u_{\mathbf{b}}-\frac{1}{2}\Phi_{\mathbf{abc}}u_{\mathbf{b}}u_{\mathbf{c}}-\cdots.
	\end{equation}
The determination of IFCs can be rewritten as a linear problem, 
\begin{equation}
\label{eq:Force}
\mathbf{F}=\mathbb{A} \mathbf{\Phi},
\end{equation}
if the unknown IFCs are arranged in a column vector $\mathbf{\Phi}$ and $\mathbf{F}$ is a column vector of the calculated atomic forces on individual atoms in a set of training configurations with displaced atoms. $\mathbb{A}$ is a matrix formed by the products of the atomic displacements in the chosen set of training configurations and is commonly referred to as the sensing matrix
	\begin{equation}
	 \mathbb{A} = \begin{bmatrix} -1 & -u_{\mathbf{b}}^1  & - \frac{1}{2}  u_{\mathbf{b}}^1 u_{\mathbf{c}}^1 & \cdots \\
                                                   & \cdots & & \\
                                                -1 & -u_{\mathbf{b}}^L  & - \frac{1}{2}  u_{\mathbf{b}}^L u_{\mathbf{c}}^L & \cdots
                                             \end{bmatrix}.
	\end{equation}
The superscripts represent a combined index labeling different atoms in the supercells of the selected training configurations; here we use one supercell with $M$ atoms and generate $N_\text{conf}$ displacement configurations, so that $L=MN_\text{conf}$. Derivative commutativity, space group symmetry and translational symmetry are used to further reduce the number of independent parameters in $\mathbf{\Phi}$. Since each symmetry operation can be written as a linear constraint on $\mathbf{\Phi}$ as
	\begin{equation}
	\mathbb{B}\mathbf{\Phi}=\mathbf{0}
	\end{equation}
and these physical constraints can be strictly enforced by finding the basis vectors of the null-space ($\mathbb{C}$) of matrix $\mathbb{B}$. If the null-space dimension of $\mathbb{B}$ is $N_{\bm{\phi}}$, we are left with $N_{\bm{\phi}}$ independent parameters in the resultant $\bm{\phi}$. The original $\mathbf{\Phi}$ can be expressed as the product of  $\mathbb{C}$ and $\bm{\phi}$
	\begin{equation}
	\mathbf{\Phi} = \mathbb{C} \bm{\phi}
	\end{equation}
where $ \mathbb{C}$ is a  $N_{\mathbf{\Phi}} \times N_{\bm{\phi}}$ matrix.

Mathematically, the solution of Eq.~\eqref{eq:Force} is obtained from a convex optimization problem that minimizes a weighted sum of the root-mean-square fitting error and the $\ell_{1}$ norm of the IFC vector $\mathbf{\Phi}$
	\begin{equation}
	\begin{split}
	\mathbf{\Phi}^\text{CS} &= {\arg \min}_\mathbf{\Phi} \; \mu \| \mathbf{\Phi} \|_1 + \frac{1}{2} \| \mathbf{F} -  \mathbb{A} \mathbf{ \Phi} \|^2_2  \\
	&\equiv {\arg \min}_\mathbf{\Phi} \; \mu \sum_{I} | \Phi_{I} | + \frac{1}{2} \sum_{a \alpha} \left( F_{a\alpha} - \sum_J A_{a\alpha, J} \Phi_J \right)^2,
	\end{split}
	\label{eq:L1min}
	\end{equation}
where $I$ is a composite index representing the collection of atomic sites and Cartesian indices defining the IFC tensor elements, $I \equiv (\mathbf{a},\mathbf{b}, \ldots )$ and $\mu$ is a parameter that adjusts the relative weights of the fitting error versus the absolute magnitude of the nonzero IFC components represented by the term with the $\ell_{1}$ norm; large values of $\mu$ favor solutions with very few nonzero IFCs at the expense of the accuracy of the fitted forces (``underfitting''), while very small $\mu$ will give a dense solution with many large nonzero IFCs that fits the DFT forces well, but may have poor predictive accuracy due to numerical noise, both random and systematic (``overfitting''). There is an optimal range of $\mu$ values between these two extremes where a sparse IFC vector $\mathbf{\Phi}$ can be obtained with excellent predictive accuracy. We note that the addition of the $\ell_1$ term solves both difficulties of the least-squares fitting approach described above: the linear problem Eq.~\eqref{eq:Force} can be underdetermined and the ill-conditioned nature of the sensing matrix does not present a problem because the $\ell_1$ norm suppresses the numerical noise stemming from the small eigenvalues of $\mathbb{A}^T\mathbb{A}$. As a consequence, according to tests performed on several systems \cite{Jiangang2016, Jiangang2017}, we find the introduction of $l_{1}$ norm term greatly improves the convergence of fitted higher-order IFCs. For example, it generally requires to calculate hundreds of supercell structures to extract 3$^{\text{rd}}$-order IFCs with a reasonable diameter cutoff, while only tens of supercell structures need to be evaluated within the CSLD framework, which significantly reduce computational cost without sacrificing accuracy. Further details of the CSLD method, numerical approaches for solving the minimization problem and its performance for calculating the lattice thermal conductivities of strongly anharmonic crystals can be found in Ref.\ [\onlinecite{csld},\onlinecite{Nelson2013},\onlinecite{Nelson20132}].

\newpage
\subsection{Phonon renormalization: Temperature-dependent atomic displacements and self-consistency }

\begin{figure}[htp]
	\includegraphics[width = 0.75\linewidth]{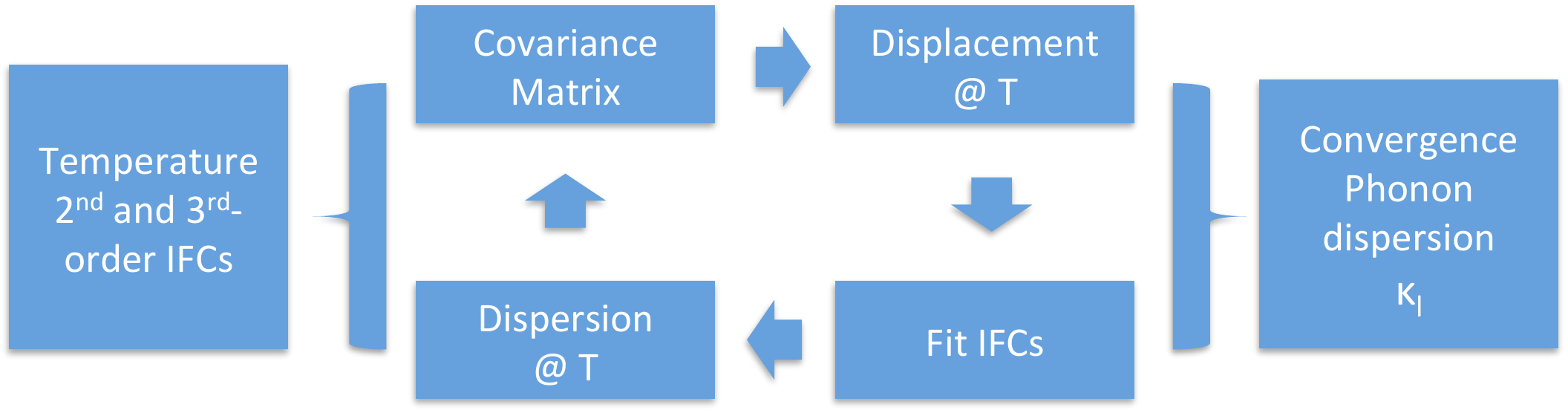}
	\caption{ 
	A schematic of iterative phonon renormalization scheme consisting of the following steps: (1) Use CSLD to obtain the irreducible representation of interatomic force constants (IFCs) and fit initial IFCs using small random displacement; (2) Evaluate quantum covariance matrix of displacement vector and generate atomic displacements according to a given temperature; (3) Calculate atomic forces of supercell structures with displaced atoms and refit IFCs using CSLD; (4) Repeat step (2) and (3) until vibrational fee energy converges within a given threshold, and compute lattice dynamics and thermal transport properties. In the above renormalization procedure, whenever an imaginary phonon frequency is encountered it is replaced with its absolute value.
	}
	\label{fig:SelfCon}
\end{figure}

To generate atomic displacements $\mathbf{u} \equiv \{ u_{a,\alpha} \}$ according to a given covariance matrix $\Sigma(\mathbf{u}) $, a lower triangle matrix $L$ needs to be computed through Cholesky decomposition of $\Sigma$, i.e., $LL^{\text{T}}=\Sigma(\mathbf{u})$. Provided that a vector $\mathbf{x}$ with the same dimension of $\mathbf{u}$ has a covariance matrix of $\Sigma(\mathbf{x})$ and the variance of $\mathbf{u}\equiv L\mathbf{x}$ is $L\text{var}(\mathbf{x})$, the covariance of $L\mathbf{x}$ becomes $L\Sigma(\mathbf{x})L^{\text{T}}$. Therefore, $\Sigma(\mathbf{u}) $ can be recovered if $\mathbf{x}$ has identity covariance matrix, indicating that $\mathbf{u}$ can be generated using $\mathbf{x}$ following either continuous Gaussian distribution or discrete Rademacher distribution. According to our tests on SrTiO$_{3}$ and VO$_{2}$, these two distributions yield very similar results if atomic configurations are sufficiently sampled, and the later generally requires fewer samples (on the order of 10 for each iteration within the CSLD framework) due to a more constrained distribution. Therefore we used discrete Rademacher distribution, while Roekeghem \textit{et.\ al.} used continuous Gaussian distribution \cite{Roekeghem2016}. Fig.\ \ref{fig:SelfCon} shows the self-consistent loop in the iterative phonon renormalization procedure. To suppress the fluctuations of computed vibrational free energy in each iteration and achieve faster convergence, it is important to introduce a mixing parameter $\eta$, which is 0.5 in this study  and used to average the phonon mode frequencies from current and previous iteration.

\newpage
\section{Computational details} 
Vienna $Ab\ initio$ Simulation Package (VASP) was used to perform structural relaxation and self-consistent calculations \cite{Vasp1, Vasp2, Vasp3, Vasp4}. The projector-augmented wave (PAW) \cite{paw} method was used in conjunction with the generalized gradient approximation (GGA) \cite{gga} for the exchange-correlation (xc) functional \cite{dft}. Perdew-Burke-Ernzerhof (PBE) and its revised version for solids (PBEsol) \cite{pbe,Perdew2008} xc functionals were used, since recent studies \cite{Budai2014,Ropo2008} show that (1) PBE tends to predict lattice dynamics properties in good agreement with experiments and (2) PBEsol generally yields better lattice parameter via achieving a more balanced binding energy compared to PBE and LDA. Experimental lattice parameters were found \cite{Budai2014}, as well as confirmed in this study (see results and discussions section), to better describe lattice dynamics properties compared to DFT-relaxed structures, therefore structures of experimental lattice parameters with relaxed internal atomic coordinates were used in subsequent calculations.

The lattice parameters of R-VO$_2$ at 425 and 600 K were obtained by interpolating experimental values \cite{McWhan1974}, and the lattice parameters of M-VO$_{2}$ at room temperature were used \cite{Longo1970}. DFT calculations of primitive cells of R (M)-VO$_{2}$ were performed using a 6$\times$6$\times$10 (6$\times$6$\times$6) $\Gamma$-centered $\mathbf{k}$-point mesh and a kinetic energy cutoff of 500 eV. The force and energy convergence thresholds were set to be $10^{-3}$ eV/\r{A} and $10^{-8}$ eV respectively. 2$\times$2$\times$4 (2$\times$2$\times$2) supercell structures were constructed,  which were sampled with 3$\times$3$\times$3 $\Gamma$-centered $\mathbf{k}$-point mesh, to extract 2$^{\text{nd}}$- and 3$^{\text{rd}}$-order IFCs for R (M)-VO$_{2}$. No explicit cutoff is imposed for 2$^{\text{nd}}$-order IFCs by taking into account the cumulant IFCs \cite{Parlinski1997}. The diameter cutoff of 3$^{\text{rd}}$-order interactions was varied from 3.0 to 4.5 \r{A} for both M- and R-VO$_{2}$ (see more details in the results and discussions section). We found that 4.0 \r{A} is sufficient to achieve good convergence of computed $\kappa_{l}$ well within 5\%. We only performed phonon renormalization for R-VO$_{2}$ because both DFT calculation and experimental measurements have confirmed that M-VO$_{2}$ mostly exhibits harmonic behavior \cite{Budai2014,Lee371}. To extract 3$^{\text{rd}}$-order IFCs of M-VO$_{2}$, 1332 2$\times$2$\times$2 supercell structures were evaluated due to low symmetry of M-VO$_{2}$ ($P2_{1}/c$). For R-VO$_{2}$, twenty 2$\times$2$\times$4 supercell structures were sampled in each iteration during phonon renormalization process, owing to its relatively high symmetry of $P4_{2}/mnm$. The ShengBTE package \cite{shengbte}, which was modified to work with CSLD, was used to perform all phonon-related calculations including the iterative calculation of $\kappa_{l}$ with a 16$\times$16$\times$26$\ ($14$\times$14$\times$14$)\ \mathbf{q}$-point mesh for R (M)-VO$_{2}$. The $\kappa_{l}$ of polycrystalline VO$_{2}$ was computed by averaging over $\kappa_{l}$ along three Cartesian coordinates.

DFT calculations of primitive cells of SrTiO$_{3}$ were performed using a 8$\times$8$\times$8 $\Gamma$-centered $\mathbf{k}$-point mesh and a kinetic energy cutoff of 520 eV. The force and energy convergence threshold were set to be $10^{-3}$ eV/\r{A} and $10^{-8}$ eV respectively. 2$\times$2$\times$2 supercell structures were constructed,  which was sampled with 4$\times$4$\times$4 $\Gamma$-centered $\mathbf{k}$-point mesh, to extract 2$^{\text{nd}}$- and 3$^{\text{rd}}$-order IFCs for SrTiO$_{3}$. In the phonon renormalization process, ten 2$\times$2$\times$2 supercell structures in each iteration were found sufficient to converge the IFCs fitting because of high symmetry of cubic SrTiO$_{3}$ ($Pm\bar{3}m$). The diameter cutoff of 3$^{\text{rd}}$-order interactions in SrTiO$_{3}$ was 5.0 \r{A}, and $\kappa_{l}$ was computed using a 20$\times$20$\times$20\ $\mathbf{q}$-point mesh.

\newpage
\section{Results and discussions}

\subsection{SrTiO$_{3}$: Validation of proposed phonon renormalization scheme}

\begin{figure}[htp]
	\includegraphics[width = 1.0\linewidth]{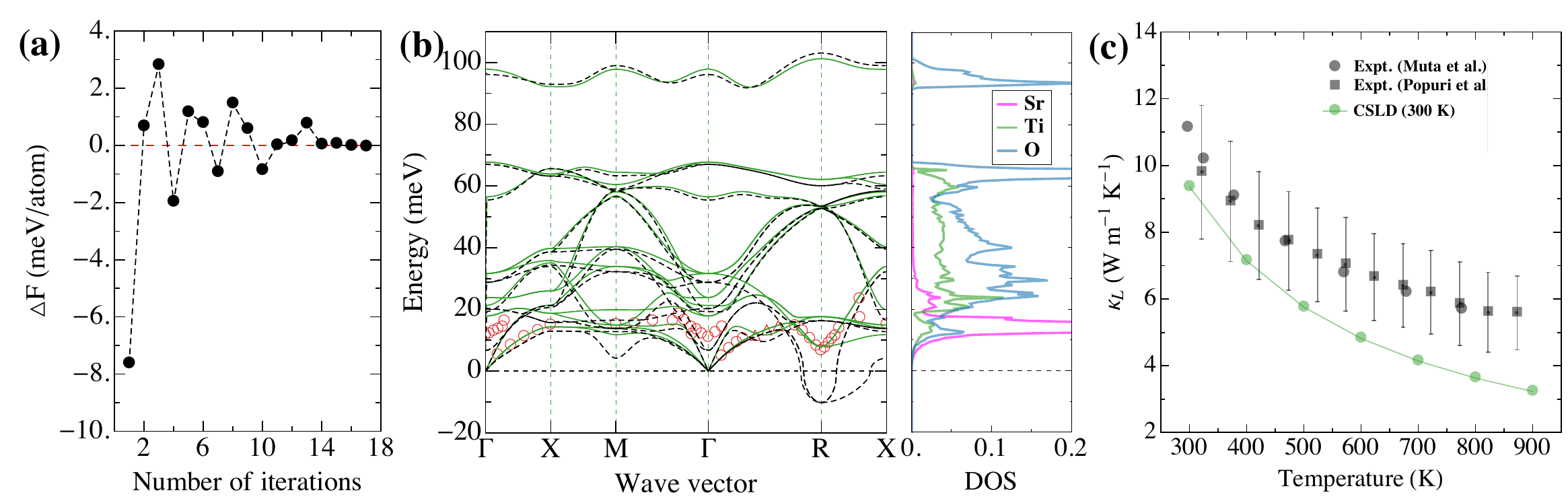}
	\caption{ 
	(a) The change of vibrational free energy between consecutive iterations as a function of the number of iterations in the phonon renormalization process of perovskite SrTiO$_{3}$ at 300 K. (b) Renormalized phonon dispersions (solid green lines) at 300 K in comparison with phonon dispersions computed in the harmonic approximation at 0 K (dashed black lines) and experimental measurements (empty triangles and circles) \cite{Stirling1972,COWLEY1969181}. The right panel shows the atom-projected phonon density of states (DOS). (c) Computed lattice thermal conductivity using renormalized interatomic force constants at 300 K in comparison with experimental measurements \cite{MUTA2005306,Popuri2014}.
	}
	\label{fig:STO}
\end{figure}

To confirm the validity of our approach and verify the implementation, we applied our phonon renormalization scheme to a well studied system: perovskite SrTiO$_{3}$ with cubic symmetry at high-temperature, which undergoes a cubic-to-tetragonal phase transition at 105 K upon cooling \cite{Cowley1964}. Fig.~\ref{fig:STO}(a) shows that the convergence of vibrational free energy within $\pm$1 meV/atom can be achieved after eleven iterations at 300 K. Additional iterations were performed to obtain forces required to converge the IFCs fitting. The phonon dispersions computed in the harmonic approximation at 0 K exhibits instability at $R$ point in the Brillouin zone of cubic cell, which corresponds to the rotation of octahedra formed by TiO$_{3}$. The renormalized phonon dispersions at 300 K, as shown in Fig.~\ref{fig:STO}(b), display well-defined phonon frequencies without imaginary numbers, further confirming that the cubic structure can be stabilized by temperature. The computed renormalized phonon dispersions at 300 K agree well with experimental measurements \cite{Stirling1972,COWLEY1969181}, particularly near the $R$ point. Using the extracted renormalized IFCs at 300 K, we also computed the $\kappa_{l}$ as a function of temperature. Fig.~\ref{fig:STO}(c) shows that the computed $\kappa_{l}$ agrees well with experiments at 300 K \cite{MUTA2005306,Popuri2014}, but shows significant underestimation at high temperatures, which is due to the fact that high-temperature IFCs are different from those at 300 K. Recently, we also applied our phonon renormalization scheme to other perovskites CsPbCl$_{3}$ and CsPbBr$_{3}$, the details of of which can be found in Ref.\ [\onlinecite{Peijun2017}]

\newpage
\subsection{VO$_{2}$: Crystal structures}
\begin{figure}[htp]
	\includegraphics[width = 1.0\linewidth]{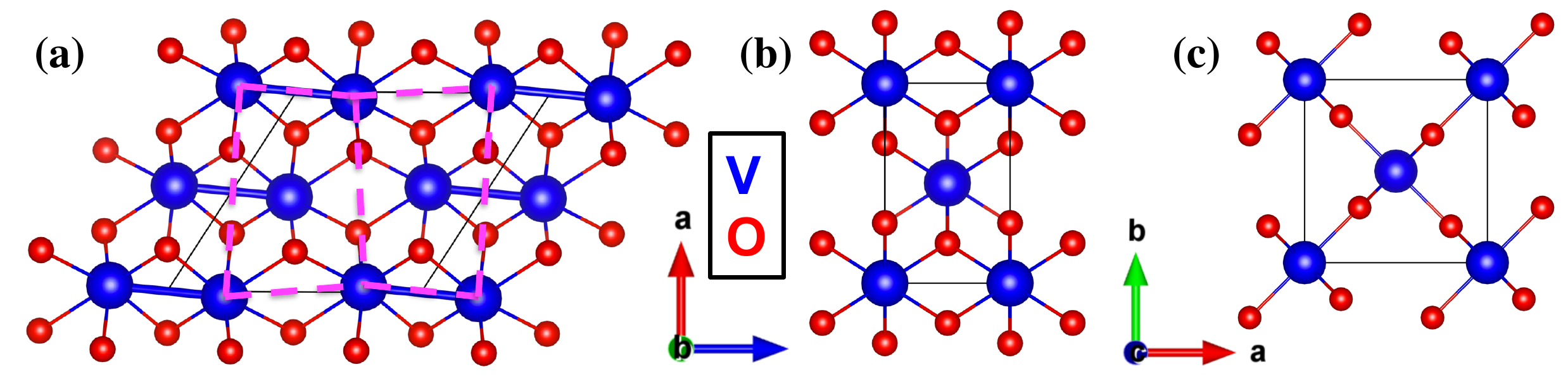}
	\caption{
	Crystal structures of (a) M-VO$_{2}$ and (b) R-VO$_{2}$ projected onto rutile $b$ axis. The dashed magenta lines denote the distorted rectangle formed by V atoms compared to the R-VO$_{2}$. The solid blue lines in (a) shows the dimerization of V atoms in monoclinic phase. (c)  Crystal structure of R-VO$_{2}$ projected onto rutile $c$ axis. Solid blue and red balls denote V and O atoms respectively.
	}
	\label{fig:Structures}
\end{figure}

R-VO$_{2}$ has a symmetry of $P4_{2}/mnm$ with one symmetrically-inequivalent V and O atom each. M-VO$_{2}$ has a symmetry of $P2_{1}/c$ with one  symmetrically-inequivalent  V and two  symmetrically-inequivalent O atoms. As shown in Fig.~\ref{fig:Structures}(a) and (b), V atoms in both phases are arranged in 1D chains along the rutile $c$ axis with O atoms octahedrally bonded to V atoms. Upon cooling, the tetragonal R-VO$_{2}$ transforms into monoclinic M-VO$_{2}$, which features zigzag dimerization of all V atoms, tilting of octahedra and doubling of the unit cell in the chain direction.

\newpage
\subsection{VO$_{2}$: Convergence of $\kappa_{l}$ with respect to the diameter cutoff of 3$^{\text{rd}}$-order interactions}
\begin{figure}[htp]
	\includegraphics[width = 0.5\linewidth]{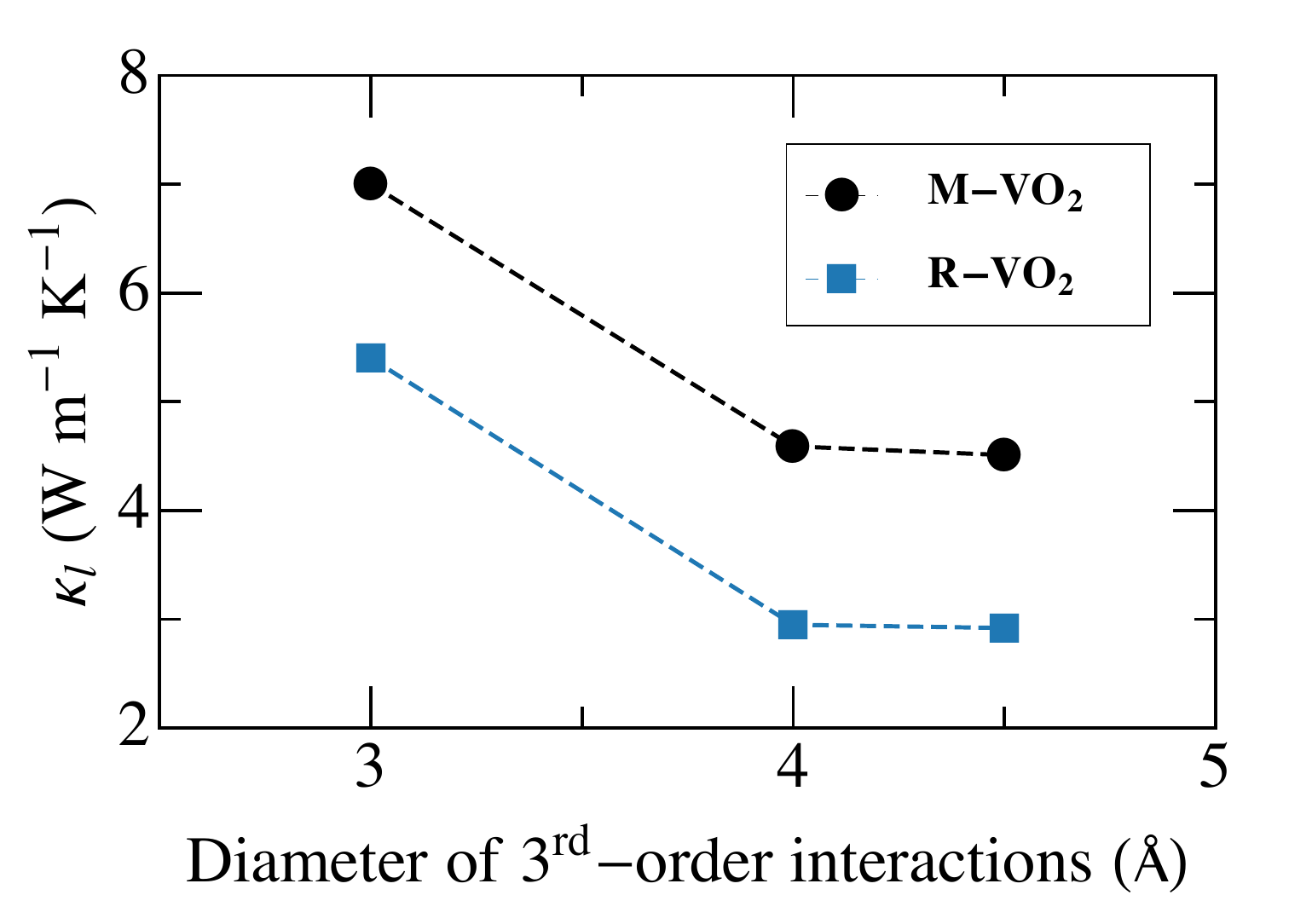}
	\caption{
	Computed lattice thermal conductivity as a function of diameter cutoff of 3$^{\text{rd}}$-order interatomic force constants.
	}
	\label{fig:Con_Kappa}
\end{figure}

Fig.~\ref{fig:Con_Kappa} shows the computed $\kappa_{l}$ as a function of diameter cutoff of 3$^{\text{rd}}$-order interactions for M-VO$_{2}$ at 340 K and R-VO$_{2}$ at 425 K respectively. This plot shows that including 3$^{\text{rd}}$-order interactions up to 4.0 \r{A} is sufficient to converge the $\kappa_{l}$.


\newpage
\subsection{VO$_{2}$: Effects of exchange correlation functional and structure relaxation on phonon dispersion}
\begin{figure}[htp]
	\includegraphics[width = 1.0\linewidth]{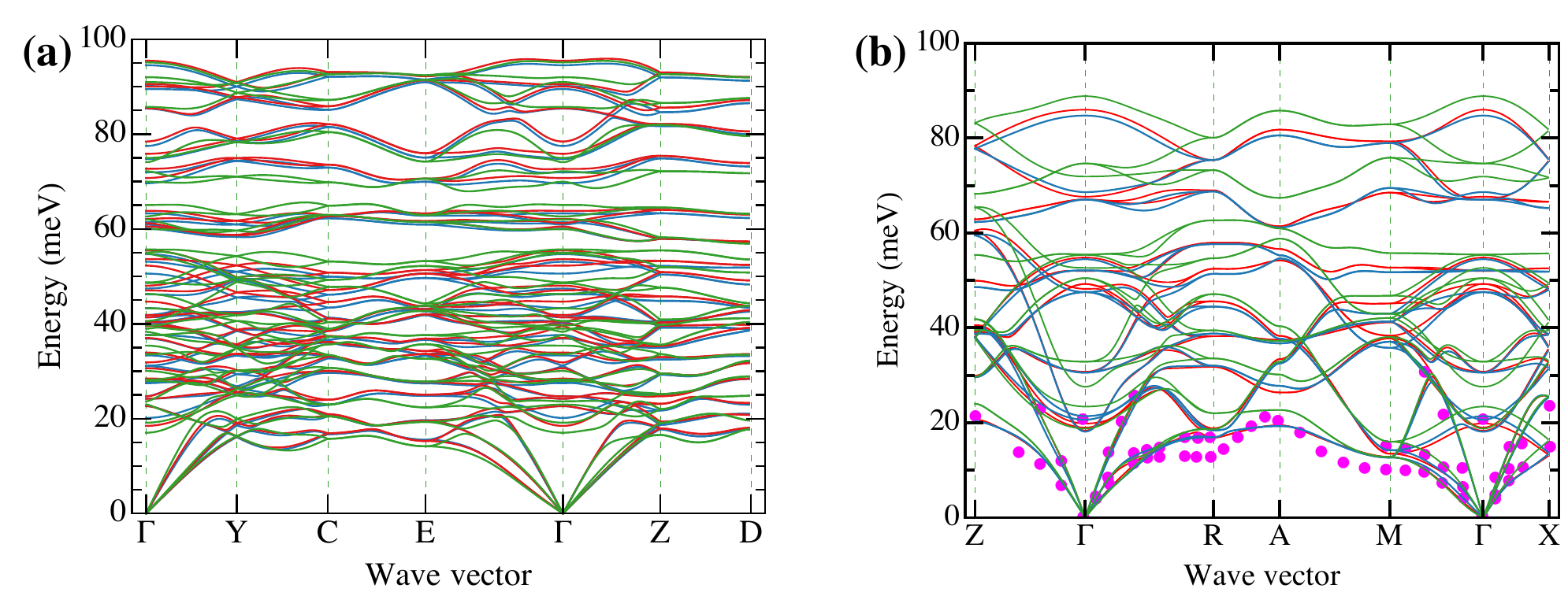}
	\caption{
	(a) and (b) Comparisons of phonon dispersions using \{PBEsol exchange correlation functional, experimental lattice parameter\} (solid blue lines), \{PBE exchange correlation functional, experimental lattice parameter\} (solid red lines) and \{PBEsol exchange correlation functional, DFT-relaxed lattice parameter\} (solid green lines) for monoclinic (a) and rutile (b) VO$_{2}$. Experimental measurements are denoted as magenta disks \cite{Budai2014}.
	}
	\label{fig:PBE_Compare}
\end{figure}

Fig.~\ref{fig:PBE_Compare}(a) shows that results obtained using PBE and PBEsol with experimental lattice parameter yield very similar phonon dispersions for M-VO$_{2}$, while PBEsol with volume-relaxed structure leads to softened phonon dispersions, which leads to values of the $\kappa_{l}$ about half of what is determined using experimental lattice parameters. Fig.~\ref{fig:PBE_Compare}(b) shows that results obtained using PBE and PBEsol with experimental lattice parameter display similar phonon dispersions for R-VO$_{2}$, while PBEsol with volume-relaxed structure leads to significantly hardened phonon dispersions, which gives rise to significantly enhanced $\kappa_{l}$, even higher than M-VO$_{2}$. We used PBE and PBEsol exchange correlation functionals together with experimental lattice parameters for both M- and R-VO$_{2}$.

\newpage
\subsection{VO$_{2}$: Renormalized phonon dispersions of R-VO$_{2}$ at 360, 425 and 600 K}
\begin{figure}[htp]
	\includegraphics[width = 1.0\linewidth]{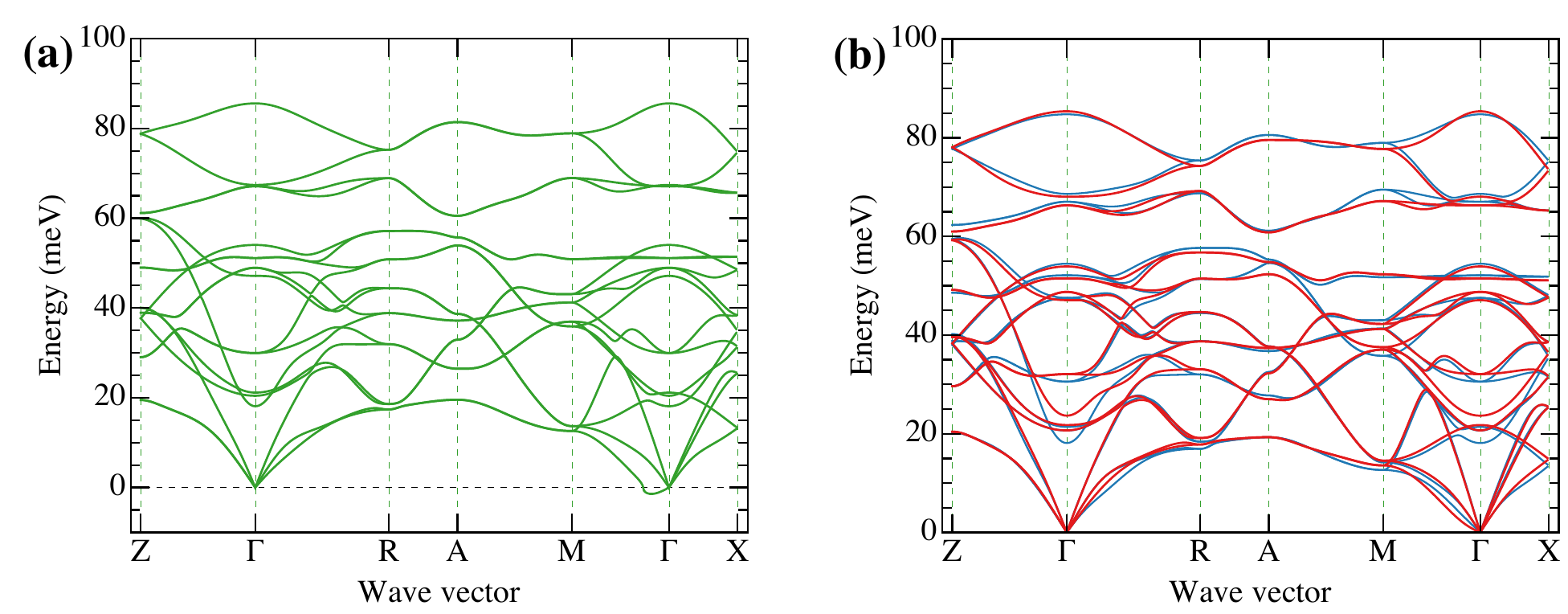}
	\caption{
	(a) Renormalized phonon dispersions of R-VO$_{2}$ at 360 K (solid green lines). (b) Comparison of renormalized phonon dispersions of R-VO$_{2}$ at 425 (solid blue lines) and 600 K (solid red lines). PBEsol exchange correlation functional and experimental lattice parameters were used.
	}
	\label{fig:Disp_Temp}
\end{figure}

Fig.~\ref{fig:Disp_Temp}(a) shows the renormalized phonon dispersions of R-VO$_{2}$ at 360 K with small lattice instabilities near $\Gamma$ point. The instability is not associated with either the R-point transverse acoustic phonon mode or the low-lying zone-center optical mode, consistent with previous theoretical investigations \cite{Budai2014}. Fig.~\ref{fig:Disp_Temp}(b) shows the comparison of renormalized phonon dispersions of R-VO$_{2}$ at 425 and 600, which show that low-lying zone-center optical mode has a strong temperature dependence.

\newpage
\subsection{VO$_{2}$: Comparisons of energy cumulative $\kappa_{l}$, phonon lifetime, group velocity, and mean free path between M-VO$_{2}$ at 340 K and R-VO$_{2}$ at 425 K.}
\begin{figure*}[htp]
	\includegraphics[width = 0.45\linewidth]{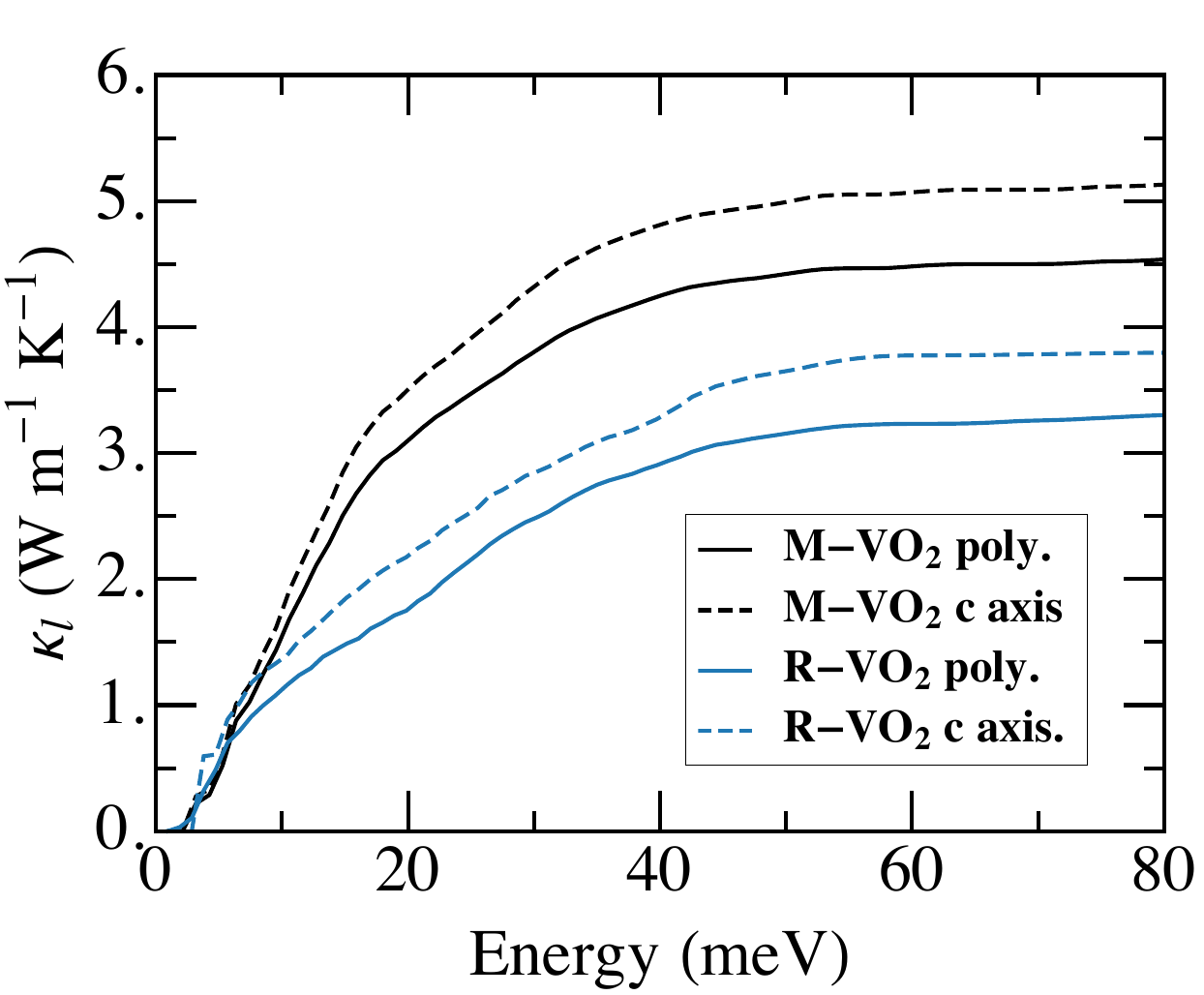}
	\caption{
	 Energy cumulative lattice thermal conductivities of poly crystal VO$_{2}$ and single crystal VO$_{2}$ along rutile $c$ axis.
	 }
	\label{fig:Kappa_Cum}
\end{figure*}

\begin{figure*}[htp]
	\includegraphics[width = 1.0\linewidth]{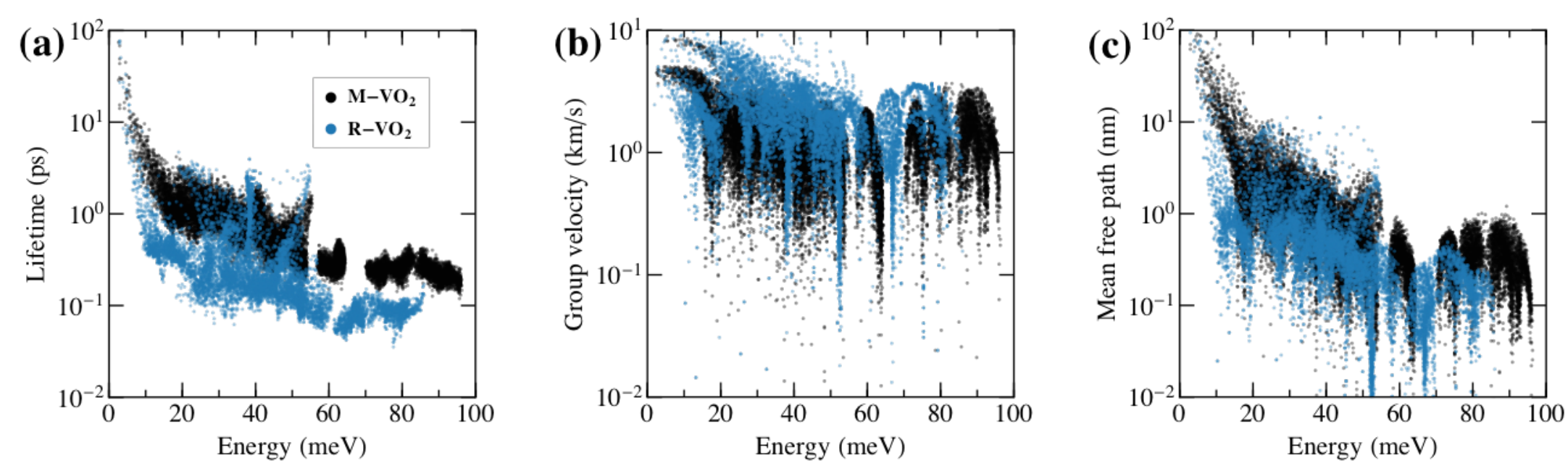}
	\caption{ (a), (b) and (c) Comparisons of phonon mode lifetime, group velocity and mean free path between monoclinic VO$_{2}$ at 340 K and rutile VO$_{2}$ at 425 K.
	 }
	\label{fig:Kappa_Ana_Sup}
\end{figure*}

Fig.~\ref{fig:Kappa_Cum} displays the energy cumulative lattice thermal conductivities of polycrystalline VO$_{2}$ and single-crystalline VO$_{2}$ along rutile $c$ axis, which indicates that phonon modes with energies less than 10 meV contribute only about 30\% of total $\kappa_{l}$.  Fig.~\ref{fig:Kappa_Ana_Sup}(a)-(c) compare the phonon mode lifetime, group velocity and mean free path between monoclinic VO$_{2}$ at 340 K and rutile VO$_{2}$ at 425 K.

\clearpage
\newpage

\end{document}